%% file: ris_based_full_duplex_Transciver_design.tex
\documentclass[conference]{IEEEtran}

\makeatletter

\newcommand{\Rmnum}[1]{\expandafter\@slowromancap\romannumeral #1@}
\makeatother
\usepackage{array}
\usepackage{multirow}
\usepackage{balance}
\usepackage{diagbox}
\usepackage{amsthm,amsmath,amsfonts,syntonly,setspace}
\usepackage{amssymb}
\usepackage{epsfig} 
\usepackage{epstopdf}
\usepackage{epsf,graphics,graphicx}
\usepackage{cite,bm,color,url,textcomp}
\usepackage{float}
\usepackage{algorithm}
\usepackage{algorithmic}
\usepackage{mathrsfs}
\usepackage{arydshln}
\usepackage{stfloats}
\usepackage{subfigure}
\usepackage{lipsum}
\usepackage{cuted}
\newcommand\blfootnote[1]{%
  \begingroup
  \renewcommand\thefootnote{}\footnote{#1}%
  \addtocounter{footnote}{-1}%
  \endgroup
}

\begin{document}
\include{com}
\title{SIMRP: Self-Interference Mitigation Using RIS and Phase Shifter Network}

\author{Wei Zhang$^1$, Ding Chen$^1$, Bin Zhou$^1$, Yi Jiang$^2$, Zhiyong Bu$^1$ \\
$^1$Shanghai Institute of Microsystem and Information Technology, Chinese Academy of Sciences, Shanghai, China\\
Email: {wzhang@mail.sim.ac.cn}, \ {chending2001@163.com}, \{bin.zhou, zhiyong.bu \}@mail.sim.ac.cn\\
$^2$School of Information Science and Technology, Fudan University, Shanghai, China \\
Email: {yijiang@fudan.edu.cn}
}


\maketitle
\blfootnote{
Work in this paper was supported by Shanghai Post-doctoral Excellence Program Grant No. 2023689.
}
\begin{abstract}
Strong self-interference due to the co-located transmitter is the bottleneck for implementing an in-band full-duplex (IBFD) system. If not adequately mitigated, the strong interference can saturate the receiver's analog-digital converters (ADCs) and hence void the digital processing. This paper considers utilizing a reconfigurable intelligent surface (RIS), together with a receiving (Rx) phase shifter network (PSN), to mitigate the strong self-interference through jointly optimizing their phases. This method, named self-interference mitigation using RIS and PSN (SIMRP), can suppress self-interference to avoid ADC saturation effectively and therefore improve the sum rate performance of communication systems, as verified by the simulation studies.
\end{abstract}

%
\IEEEpeerreviewmaketitle

\section{Introduction}
Through transmitting (Tx) and receiving (Rx) signals in the same frequency band simultaneously, an in-band full-duplex (IBFD) system can achieve twice the capacity of its counterpart working in either a time division duplex (TDD) mode or a frequency division duplex (FDD) mode. In practice, however, the strong self-interference, caused by the co-located transmitter, can saturate the receiver's analog-digital converters (ADCs), and hence void the subsequent digital signal processing of the co-located receiver \cite{6832464}. Therefore, it is a prerequisite to mitigate the strong self-interference before the ADCs for implementing a real IBFD system.

In the past decades, numerous approaches have been proposed for self-interference mitigation (SIM). Initially, the papers \cite{101145} adopted symmetric placement of antennas to mitigate self-interference, but only for narrowband signals. Some researchers proposed to create a copy of self-interference and subtract it from the received signal before the ADCs, where the copy can be reconstructed from the digital domain \cite{9153162} or the radio frequency (RF) domain \cite{10160119}. However, due to hardware overheads, these analog SIM techniques do not apply to multi-input multi-output (MIMO) systems. Hence the paper \cite{SoftNull} proposed to optimize the Tx beamforming matrix to null the effective channel of the strong self-interference for MIMO systems. Moreover, the emergence of reconfigurable intelligent surfaces (RISs) has opened up new possibilities for SIM. For example, the paper \cite{9839223} utilized RIS to mitigate the self-interference with a Rx antenna and a Tx antenna and built a prototype that can achieve the SIM amount of $85$dB. The paper \cite{10153667} proposed to optimize the Tx beamforming matrix and RIS coefficients for SIM of a RIS-assisted IBFD (RAIBFD) system, which has over $80\%$ gain of sum rate compared with the traditional FDD system, but utilizing Tx beamforming matrix for SIM also reduces the downlink (DL) capacity \cite[Fig. 6]{10153667}, which is undesirable as the transmission demand of DL data is much stronger than that of uplink in cellular networks.

Analog phase shifter network (PSN), deployed in between the Rx antennas and the ADCs, shows great compatibility with the MIMO systems, making it a perfect tool to mitigate the strong self-interference for future IBFD systems. In \cite{10210613}, the authors utilized RIS and PSN to enhance the spectral efficiency for full-duplex mmWave MIMO systems, but it did not consider the effect of finite bit resolution of ADC, and therein the SIM is not conducted before the ADCs.

This paper considers a RIS-assisted IBFD transceiver equipped with a PSN at Rx RF chains, and we propose to jointly optimize the phases of RIS and PSN to null the effective channel of the self-interference before the Rx ADCs, which is called self-interference mitigation using RIS and PSN (SIMRP). Compared with the RAIBFD system proposed in \cite{10153667}, the proposed SIMRP can release the downlink capacity potential and thus increase the sum rate of the IBFD system, as verified in the simulation results where the proposed SIMRP has a $33\%$ gain of sum rate over the RAIBFD, $70\%$ gain of sum rate over the conventional RIS-assisted FDD (RAFDD) system, and also reaches $95\%$ of the sum rate of the ideal IBFD system equipped with $\infty$-bit ADCs.

\section{Signal Model and Problem Formulation} \label{SEC2}

\subsection{Signal Model}
\begin{figure}[htb]
\centering
{\psfig{figure=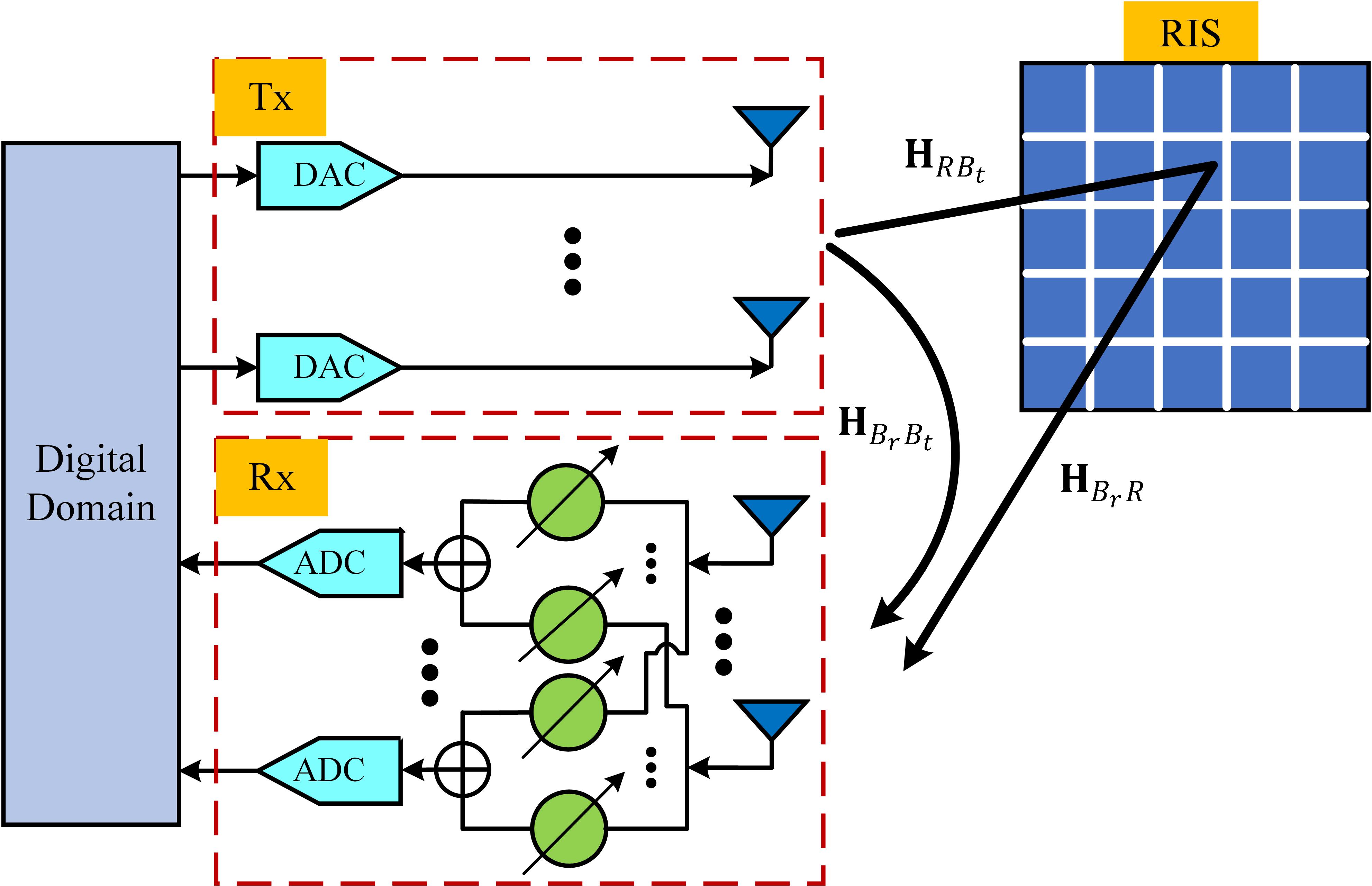,width= 3.5in}}
\caption{The proposed IBFD transceiver structure.}
\label{fig.Framework}
\end{figure}

We consider the self-interference mitigation (SIM) for an in-band full-duplex (IBFD) transceiver aided by a co-located $M_{ris}$-element RIS as shown in Fig. \ref{fig.Framework}. The transceiver consists of $M_t$ transmitting (Tx) antennas, $M_r$ receiving (Rx) antennas, and $N_{\rm RF}$ Rx radio frequency (RF) chains. An analog phase shifter network (PSN), denoted as $\Fbf_{\rm RF}\in \Scal^{N_{\rm RF}\times M_{r}}$ with $\Scal = \{e^{j\theta}|\theta\in\Rnum\}$ and $N_{\rm RF} \le M_r$, is deployed in between the Rx antennas and the ADCs. Denote the channel from the Tx antennas to the Rx antennas as  $\Hbf_{B_rB_t}\in{\mathbb C}^{M_r\times M_t}$, the channel from the Tx antennas to the RIS as $\Hbf_{RB_t}\in{\mathbb C}^{M_{ris}\times M_t}$ and the channel from the RIS to the Rx antennas as  $\Hbf_{B_rR}\in{\mathbb C}^{M_r\times M_{ris}}$. Since the RIS channel estimation techniques have been widely investigated \cite{9133156}\cite{10313265}, we assume the channel state information (CSI) to be perfectly known in the rest of this paper.

As the Tx antennas emit signal $\sbf\in{\mathbb C}^{M_{t}\times 1}$, the self-interference at the output of the PSN (but before the ADCs) can be denoted as
\ben
\xibf = \Fbf_{\rm RF}(\Hbf_{B_rR}\Dbf\Hbf_{RB_t}+\Hbf_{B_rB_t})\sbf,
\label{equ.si}
\een
where $\Dbf = \diag\left(e^{j\phi_1},e^{j\phi_2},\dots,e^{j\phi_{M_{ris}}}\right)$ represents the adjustable phases of the RIS elements.

\subsection{Problem Formulation}
As the self-interference, along with the uplink signal, enters the Rx ADCs, the relations between the input signal-to-interference ratio (SIR) and the output signal-to-quantization-plus-noise ratio (SQNR) can be denoted as \cite{9420257}
\ben
{\sf SQNR}_{\rm dB} \approx {\sf SIR}_{\rm dB} + 6.02{\sf ENOB}-4.35,
\label{equ.qNPow}
\een
where ENOB represents the effective number of bits of ADCs. According to (\ref{equ.qNPow}), a self-interference of $30$dBm  may lead to quantization noise of $30-72+4.35=-37.65$dBm given that ENOB $= 12$, which shows that the self-interference mitigation (SIM) should be conducted before the Rx ADCs. To this end, we propose a SIM method named self-interference mitigation using RIS and PSN (SIMRP), which aims to jointly optimize the RIS coefficients $\Dbf$ and the PSN $\Fbf_{\rm RF}$ to null the effective channel of the self-interference, i.e., to solve
\begin{subequations}
\begin{align}
\mathop{\min}_{\Dbf,\Fbf_{\rm RF}} \ &\|\Fbf_{\rm RF}(\Hbf_{B_rR}\Dbf\Hbf_{RB_t}+\Hbf_{B_rB_t}) \|_F^2, \label{equ.objFuncOria}\\
\text{s.t.}\quad &\Dbf = \diag\left(e^{j\phi_1},e^{j\phi_2},\dots,e^{j\phi_{M_{ris}}}\right),\label{equ.objFuncOrib} \\
& \Fbf_{\rm RF}\Fbf_{\rm RF}^H = M_r\Ibf, \label{equ.objFuncOric}\\
& \Fbf_{\rm RF} \in \Scal^{N_{\rm RF}\times M_{r}},
\label{equ.objFuncOrid}
\end{align}
\label{equ.objFuncOri}
\end{subequations}
\!\!\!\!where (\ref{equ.objFuncOric}) ensures that $N_{\rm RF}$ dimensions of subspace are reserved for uplink signal transmission (cf. \cite[(4)]{SoftNull}). Given that (\ref{equ.objFuncOri}) is non-convex and the optimal solution appears intractable, we propose an alternating optimization method to obtain its near-optimal solution.

\section{Algorithm for Optimizing the Coefficients of RIS and PSN}\label{sec.RISopt}
In this section, we propose an alternating optimization method to solve (\ref{equ.objFuncOri}), and obtain a near-optimal solution of $\Dbf$ and $\Fbf_{\rm RF}$.
\subsection{The Optimization of $\Dbf$} \label{sec.dopt}
Fixing $\Fbf_{\rm RF}$, we have from (\ref{equ.objFuncOri}) that
\begin{align}
\mathop{\min}_{\Dbf} \ &  \|\Abf\Dbf\Hbf_{RB_t}+ \Bbf\|_F^2, \nonumber\\
\text{s.t.}\quad &\Dbf = \diag\left(e^{j\phi_1},e^{j\phi_2},\dots,e^{j\phi_{M_{ris}}}\right),
\label{equ.objFuncOriv2}
\end{align}
where $\Abf = \Fbf_{\rm RF}\Hbf_{B_rR} \in {\mathbb C}^{N_{\rm RF}\times M_{ris}}$ and $\Bbf = \Fbf_{\rm RF}\Hbf_{B_rB_t}\in {\mathbb C}^{N_{\rm RF}\times M_{t}}$.
Using the formula ${\rm vec}(\Xbf\Ybf\Zbf) = (\Zbf^T*\Xbf)\ybf$ where $\Ybf = \diag(\ybf)$ is a diagonal matrix with $\ybf$ being its diagonal elements and $*$ represents Khatri–Rao product, we have the equivalent form of (\ref{equ.objFuncOriv2}) as
\begin{subequations}
\begin{align}
\mathop{\min}_{\dbf} \ & f(\dbf) \triangleq \|\Cbf\dbf+ \bbf\|_2^2, \label{equ.objFuncOriv3a}\\
\text{s.t.}\quad &\dbf = \left[e^{j\phi_1},e^{j\phi_2},\dots,e^{j\phi_{M_{ris}}}\right]^T,
\label{equ.objFuncOriv3b}
\end{align}
\label{equ.objFuncOriv3}
\end{subequations}
\!\!\!where $\Cbf = \Hbf_{RB_t}^T*\Abf \in {\mathbb C}^{M_{t}N_{\rm RF}\times M_{ris}}$, $\bbf = {\rm vec}(\Bbf)$, and ${\rm vec}(\Bbf)$ stands for the column vector obtained by stacking the columns of $\Bbf$ into a single column.

As the constant modulus of the variable $\dbf$ defines a manifold
\ben
\Mcal_{cc}^{M_{ris}} \triangleq \left\{\dbf\in{\Cnum}^{M_{ris}}: |\dbf(1)| =  \cdots = |\dbf({M_{ris}})| = 1\right\},
\label{equ.mafd}
\een
we can solve (\ref{equ.objFuncOriv3}) using the Riemannian conjugate gradient (RCG) algorithm, which can obtain a suboptimal solution as follows. The Riemannian gradient of $f(\dbf)$ is defined as ${\rm grad}({\dbf}) \triangleq \nabla f({\dbf}) -{\rm Re}\{\nabla f({\dbf})\circ {\dbf}^*\}\circ {\dbf}$, where the Euclidean gradient $\nabla f({\dbf})$ is
\ben
\nabla f(\dbf) = \Cbf^H(\Cbf\dbf+\bbf),
\label{equ.EucGrad}
\een
and $\circ$ represents the Hadamard product. Starting from point ${\dbf}_i$ in the $i$-th iteration with Riemannian gradient $\gbf_{i} = {\rm grad}({\dbf}_i)$, we can reach another point as
\ben
\tbf_i = {\dbf}_{i}+\alpha_{i}\cbf_{i},
\een
where $\alpha_{i}$ is the step size obtained using the Armijo{-}Goldstein condition and $\cbf_i$ is the conjugate direction defined as
\ben
\cbf_{i} \triangleq \left\{
\ba{ll} -\gbf_{0}, & i = 0, \\
-\gbf_{i} + \beta_{i}\cbf_{i-1}^{+}, & i \ge 1.
\ea \right.
\een
$\beta_{i}\triangleq \frac{||\gbf_{i}||_2^2}{||\gbf^+_{i-1}||_2^2}$,
$\cbf_{i-1}^{+}$ and $\gbf_{i-1}^{+}$ are obtained as
\ben
\begin{split}
\cbf_{i-1}^{+} &= \cbf_{i-1} - {\rm Re}\{\cbf_{i-1}\circ \dbf_{i}^*\}\circ\dbf_{i}, i\ge 1,\\
\gbf_{i-1}^{+} &= \gbf_{i-1} - {\rm Re}\{\gbf_{i-1}\circ \dbf_{i}^*\}\circ\dbf_{i}, i\ge 1.
\end{split}
\label{equ.trasp}
\een
Then we map $\tbf_i$ onto the manifold from (\ref{equ.mafd}),  i.e.,
${\rm Retr}_{\Mcal_{cc}^{M_{ris}}}(\tbf_i) \triangleq \left[\frac{\tbf_i(1)}{|\tbf_i(1)|},\frac{\tbf_i(2)}{|\tbf_i(2)|},\dots,\frac{\tbf_i(M_{ris})}{|\tbf_i(M_{ris})|}\right]^T$, and
$\dbf_{i}$ can be updated as $\dbf_{i+1} = {\rm Retr}_{\Mcal_{cc}^{M_{ris}}}(\tbf_i)$. Iterating $i$ from $0$ to $\infty$ until $f(\dbf)$ improves less than $\epsilon_1$, e.g., $\epsilon_1 = 10^{-5}$, a near-optimal solution $\dbf$ can be obtained over the Riemannian manifold.


\subsection{The Optimization of $\Fbf_{\rm RF}$} \label{Sec.FrfOpt}
Given that $\Dbf$ is obtained, we rewrite (\ref{equ.objFuncOri}) as
\begin{subequations}
\begin{align}
\mathop{\min}_{\Fbf_{\rm RF}} \ &\|\Fbf_{\rm RF}\Gbf\|_F^2, \nonumber\\
\text{s.t.} \quad & \Fbf_{\rm RF}\Fbf_{\rm RF}^H = M_r\Ibf_{N_{\rm RF}}, \label{equ.objFuncOriv4a}\\
& \Fbf_{\rm RF} \in \Scal^{N_{\rm RF}\times M_{r}},
\label{equ.objFuncOriv4b}
\end{align}
\label{equ.objFuncOriv4}
\end{subequations}
\!\!\!where $\Gbf = \Hbf_{B_rR}\Dbf\Hbf_{RB_t}+\Hbf_{B_rB_t} \in {\mathbb C}^{M_r\times M_t}$.
Denoting $\Fbf_{\rm RF} \triangleq \left[\fbf_{1},\fbf_{2},\dots,\fbf_{N_{\rm RF}}\right]^H$,
we can divide (\ref{equ.objFuncOriv4}) into $N_{\rm RF}$ subproblems, i.e.,
\begin{align}
\mathop{\min}_{\fbf_{1}} \ & \fbf_{1}^H\Gbf\Gbf^H\fbf_{1}, \nonumber\\
\text{s.t.} \quad &  \fbf_{1} \in \Scal^{N_{\rm RF}\times 1},
\label{equ.objpsn1}
\end{align}
and
\begin{subequations}
\begin{align}
\mathop{\min}_{\fbf_{m}} \ & \fbf_{m}^H\Gbf\Gbf^H\fbf_{m}, m = 2,\dots,N_{\rm RF}, \nonumber\\
\text{s.t.}\quad & \fbf_{m}^H\Fbf_{m-1}^H\Fbf_{m-1}\fbf_{m} = 0, \label{equ.objFuncpsnv4a}\\
\quad &  \fbf_{m} \in \Scal^{N_{\rm RF}\times 1},
\label{equ.objFuncpsnv4b}
\end{align}
\label{equ.objFuncpsnv4}
\end{subequations}
\!\!\!\!where $\Fbf_{m-1} = \left[\fbf_{1},\fbf_{2},\dots,\fbf_{m-1}\right]^H\in{\mathbb C}^{(m-1)\times M_r}$.
The solution to (\ref{equ.objpsn1}) can also be obtained using the RCG algorithm, while the constraint (\ref{equ.objFuncpsnv4a}) makes (\ref{equ.objFuncpsnv4}) intractable. Thus we propose an iterative sequential quadratic programming (SQP) method based on trust-region to solve (\ref{equ.objFuncpsnv4}), which is detailed as follows.
Denoting $\Theta_m = [\theta_{m,1},\theta_{m,2},\dots,\theta_{m,M_r}]^T$ where $\theta_{m,n}$ represents the phase of the $(m,n)$-th element of $\Fbf_{\rm RF}$, we can define $g(\Theta_m) \triangleq {\fbf}_m^H\Gbf\Gbf^H{\fbf}_m$ and
$h(\Theta_m)\triangleq {\fbf}_m^H\Fbf_{m-1}^H\Fbf_{m-1}{\fbf}_m$, and thus
reformulate (\ref{equ.objFuncpsnv4}) into
\begin{subequations}
\begin{align}
\mathop{\min}_{\Theta_{m}} \ & g(\Theta_m), \label{equ.objFuncpsnv5a}\\
\text{s.t.}\quad & h(\Theta_m) = 0. \label{equ.objFuncpsnv5b}
\end{align}
\label{equ.objFuncpsnv5}
\end{subequations}
\!\!\!\!\!Given the point $\Theta_{m,k}$ in the $k$-th iteration, the Lagrange function of (\ref{equ.objFuncpsnv5}) is
$\Lcal(\Theta_{m,k}, \lambda_{m,k}) = g(\Theta_{m,k}) + \lambda_{m,k} h(\Theta_{m,k})$
with $\lambda_{m,k}$ being the Lagrange multiplier. Letting $\nabla \Lcal(\Theta_{m,k}, \lambda_{m,k}) = 0$, we have
\ben
\lambda_{m,k} = -\frac{\hbf_{m,k}^T\gbf_{m,k}}{||\hbf_{m,k}||_2^2},
\label{equ.lma}
\een
where $\gbf_{m,k}$ and $\hbf_{m,k}$ represent $\nabla g(\Theta_{m,k})$ and $\nabla h(\Theta_{m,k})$ and can be obtained as
\ben
\begin{split}
\gbf_{m,k} &= -2{\rm Im}\left((\Ebf{\fbf}_{m,k}^*)\odot{\fbf}_{m,k}\right), \\
\hbf_{m,k} & = -2{\rm Im}\left((\Jbf_m{\fbf}_{m,k}^*)\odot{\fbf}_{m,k}\right),
\end{split}
\label{equ.nablagh}
\een
with $\fbf_{m,k} = e^{j\Theta_{m,k}}$, $\Ebf = \Gbf^*\Gbf^T$, and $\Jbf_m = \Fbf_{m-1}^T\Fbf_{m-1}^*$. We update $\Theta_{m,k}$ using $\Theta_{m,k+1}=\Theta_{m,k}+\Delta\Theta_{m,k}$ where $\Delta\Theta_{m,k}$ is the variation step, and thus can further approximate (\ref{equ.objFuncpsnv5}) at point $\Theta_{m,k}$ as
\begin{subequations}
\begin{align}
\mathop{\min}_{\Delta\Theta_{m,k} } \ & \gbf_{m,k}^T\Delta\Theta_{m,k}  + \frac{1}{2}\Delta\Theta_{m,k} ^T\Qbf_{m,k}\Delta\Theta_{m,k} , \label{equ.objFuncpsnv7a}
\\
\text{s.t.}\quad & \hbf_{m,k}^T\Delta\Theta_{m,k} +h(\Theta_{m,k}) = 0, \label{equ.objFuncpsnv7b} \\
& ||\Delta\Theta_{m,k} ||_2 \le \Omega_k, \label{equ.objFuncpsnv7c}
\end{align}
\label{equ.objFuncpsnv7}
\end{subequations}
\!\!\!where $\Qbf_{m,k} = \nabla^2 g(\Theta_{m,k}) + \lambda_{m,k} \nabla^2 h(\Theta_{m,k})$;
(\ref{equ.objFuncpsnv7c}) defines a trust-region that ensures that  (\ref{equ.objFuncpsnv7a}) is accurate, and $\Omega_k > 0$ is the trust-region radius.
Letting $\Sbf_{m,k} = {\fbf}_{m,k}{\fbf}_{m,k}^H$, we have
\ben
\nabla^2 g(\Theta_{m,k}) = -2{\rm Re}\left[\diag((\Ebf{\fbf}_{m,k}^*)\odot{\fbf}_{m,k})-\Ebf\odot\Sbf_{m,k}\right],
\label{equ.nablag2}
\een
\ben
\nabla^2 h(\Theta_{m,k}) = -2{\rm Re}\left[\diag((\Jbf_m{\fbf}_{m,k}^*)\odot{\fbf}_{m,k})-\Jbf_m\odot\Sbf_{m,k}\right].
\label{equ.nablah2}
\een

Noticing that $\Qbf_{m,k}$ may not be a semi-positive definite matrix, we cannot treat (\ref{equ.objFuncpsnv7}) as a convex problem. We divide $\Delta\Theta_{m,k} $ into \cite{Gould2003}
\ben
\Delta\Theta_{m,k}  = \gamma\hbf_{m,k}+\Zbf_{m,k}\xbf,
\label{equ.dThetaDiv}
\een
where $\gamma\in \mathbb{R}$ and $\Zbf_{m,k}\in \mathbb{R}^{M_r\times (M_r-1)}$ is a matrix whose columns form a basis of the orthogonal space of $\hbf_{m,k}$ with $\Zbf_{m,k}^T\Zbf_{m,k}=\Ibf$. Hence we have $\hbf_{m,k}^T\Zbf_{m,k}\xbf = 0$, leading to
\ben
||\Delta\Theta_{m,k} ||_2^2 = \gamma^2||\hbf_{m,k}||_2^2 + ||\xbf||_2^2.
\label{equ.uv}
\een

First, we obtain $\gamma$ that satisfies (\ref{equ.objFuncpsnv7b}) and (\ref{equ.objFuncpsnv7c}). Inserting (\ref{equ.dThetaDiv}) into (\ref{equ.objFuncpsnv7b}) yields that $\gamma||\hbf_{m,k}||_2^2+h(\Theta_{m,k}) = 0$, and thus we need to solve
\begin{subequations}
\begin{align}
\mathop{\min}_{\gamma}  & \  \varphi(\gamma) \triangleq \left(\gamma||\hbf_{m,k}||_2^2+h(\Theta_{m,k})\right)^2, \label{equ.objFuncpsnv8a} \\
\text{s.t.} & \quad -{\xi\Omega_k} \le \gamma||\hbf_{m,k}||_2 \le {\xi\Omega_k}, \label{equ.objFuncpsnv8b}
\end{align}
\label{equ.objFuncpsnv8}
\end{subequations}
\!\!\!where $\xi$ is a real number and $0<\xi<1$. As $\varphi(\gamma)$ is a quadratic function about $\gamma$, the closed-form solution to (\ref{equ.objFuncpsnv8}) can be obtained as
\ben
\gamma_{opt} = \left\{
\ba{ll}
\frac{\xi\Omega_k}{{||\hbf_{m,k}||_2}}, & -\frac{h(\Theta_{m,k})}{\xi||\hbf_{m,k}||_2} \ge \Omega_k, \\
-\frac{h(\Theta_{m,k})}{||\hbf_{m,k}||_2^2}, &  -\Omega_k < -\frac{h(\Theta_{m,k})}{\xi||\hbf_{m,k}||_2}  < \Omega_k, \\
-\frac{\xi\Omega_k}{{||\hbf_{m,k}||_2}}, &  \frac{h(\Theta_{m,k})}{\xi||\hbf_{m,k}||_2} \ge \Omega_k.
\ea
\right.
\label{eq.alphaopt}
\een

Second, we aim to minimize (\ref{equ.objFuncpsnv7a}) by optimizing $\xbf$. According to (\ref{equ.dThetaDiv}) and (\ref{equ.uv}), we have
\ben
\Delta\Theta_{m,k} =  \Zbf_{m,k}\xbf + \gamma_{opt}\hbf_{m,k},
\label{equ.Zx}
\een
\ben
||\Delta\Theta_{m,k} ||_2^2 = ||\xbf||_2^2 + \gamma_{opt}^2||\hbf_{m,k}||_2^2.
\label{equ.x}
\een
Then inserting (\ref{equ.Zx}) into (\ref{equ.objFuncpsnv7a}) and (\ref{equ.x}) into (\ref{equ.objFuncpsnv7c}) yields that
\begin{subequations}
\begin{align}
\mathop{\min}_{\xbf} \ & (\gbf_{m,k}+\gamma_{opt}\Qbf_{m,k}\hbf_{m,k})^T\Zbf_{m,k}\xbf + \frac{1}{2}
      \xbf^T\bar{\Qbf}_{m,k}\xbf, \label{equ.objFuncpsnv9a}\\
\text{s.t.} \ & ||\xbf||_2 \le \sqrt{\Omega_k^2 - \gamma_{opt}^2||\hbf_{m,k}||_2^2}, \label{equ.objFuncpsnv9b}
\end{align}    \label{equ.objFuncpsnv9}
\end{subequations}
\!\!\!where $\bar{\Qbf}_{m,k} = \Zbf_{m,k}^T\Qbf_{m,k}\Zbf_{m,k}$. We can obtain the approximate solution to (\ref{equ.objFuncpsnv9}) using the truncated conjugate gradient (TCG) method \cite{TCG}, from which we can further have $\Delta\Theta_{m,k} $ and $\Theta_{m,k+1}$.
Iterating $m$ from $2$ to $N_{rf}$ and $k$ from $1$ to $\infty$ until (\ref{equ.objFuncOriv4}) improves less than $\epsilon_2$, e.g., $10^{-5}$,  we can obtain a near-optimal solution to (\ref{equ.objFuncOriv4}).

We summarize this iterative algorithm in Algorithm \ref{Algo.1}, where we reduce the radius of the trust-region in line $11$ to increase the approximation accuracy of (\ref{equ.objFuncpsnv7}) and guarantee the convergence of Algorithm \ref{Algo.1}; $\Fbf_{\rm RF}(m,:)$ in line $18$ denotes the $m$-th row of $\Fbf_{\rm RF}$.
We also summarize the proposed alternating optimization between $\Fbf_{\rm RF}$ and $\Dbf$ in Algorithm \ref{Algo.2}, where $\epsilon_3 = 10^{-5}$ in line $1$ and ${\Qcal}(\cdot)$ in line $7$ can quantize the vector element-wise to the grids $\{0,\frac{2\pi}{2^b},\frac{4\pi}{2^b},\dots,\frac{(2^b-1)2\pi}{2^b}\}$ for $b$-bit RIS coefficients.
\begin{algorithm}[tb]
\caption{The Iterative SQP Method to Solve (\ref{equ.objFuncpsnv4})}
\label{Algo.1}
\begin{algorithmic}[1]
\REQUIRE $\Gbf$, $\Omega_1$, $\xi$, $\epsilon_2$ and initial $\Theta_{1,1},\dots,\Theta_{N_{\rm RF},1
}$;
\ENSURE The matrix $\Fbf_{\rm RF}$;
\STATE Solve (\ref{equ.objpsn1}) to obtain $\fbf_1$ with the power method;
\FOR{$m$ = $2$ : $N_{\rm RF}$}
\STATE Set $k=1$;
\STATE Calculate $\gbf_{m,1}$ and $\hbf_{m,1}$ from (\ref{equ.nablagh});
\STATE Calculate $\lambda_{m,1}$ from (\ref{equ.lma}), and then obtain $\Qbf_{m,1}$ from (\ref{equ.nablah2});
\WHILE{The cost function (\ref{equ.objFuncpsnv7a}) improves less than $\epsilon_2$.}
\STATE Obtain $\alpha_{opt}$ from (\ref{eq.alphaopt}), and calculate $\bar{\Qbf}_{m,k}$;
\STATE Obtain $\xbf$ from (\ref{equ.objFuncpsnv9}) using the TCG method;
\STATE Calculate $\Delta\Theta_{m,k} $ from (\ref{equ.dThetaDiv}) with $\alpha_{opt}$ and $\xbf$;
\IF{$g({\Theta_{m,k}}+\Delta\Theta_{m,k} ) > g(\Theta_{m,k})$}
\STATE $\Omega_{k+1} = 0.5\Omega_k$;
\ELSE
\STATE ${\Theta_{m,k+1}} = {\Theta_{m,k}}+\Delta\Theta_{m,k} $;
\STATE $k=k+1$;
\STATE Calculate $\gbf_{m,k}$, $\hbf_{m,k}$, $\lambda_{m,k}$, and $\Qbf_{m,k}$;
\ENDIF
\ENDWHILE
\STATE $\fbf_m = e^{j\Theta_{m}}$, $\Fbf_{\rm RF}(m,:)=\fbf_m^H$;
\ENDFOR
\end{algorithmic}
\end{algorithm}

\begin{algorithm}[htb]
\caption{Alternating Optimization between $\Fbf_{\rm RF}$ and $\Dbf$}
\label{Algo.2}
\begin{algorithmic}[1]
\REQUIRE $\Hbf_{B_rB_t}$, $\Hbf_{B_rR}$, $\Hbf_{RB_t}$, and initial $\Fbf_{\rm RF}$ and $\Dbf$;
\ENSURE The matrix $\Dbf$ and analog PSN $\Fbf_{\rm RF}$;
\WHILE {the cost function in (\ref{equ.objFuncOri}) decreases less than $\epsilon_3$}
\STATE Obtain $\Gbf$ and perform Algorithm \ref{Algo.1} to obtain $\Fbf_{\rm RF}$;
\STATE Fix $\Fbf_{\rm RF}$ and obtain $\dbf$ using the RCG algorithm from Section \ref{sec.dopt};
\IF{$b =\infty$}
\STATE$\Dbf = \diag(\dbf)$.
\ELSE
\STATE $\dbf_q = e^{j{{\Qcal}(\angle{\dbf})}}$.
\IF{$||\Cbf\dbf_q+\bbf||_2^2<||\Cbf\dbf+\bbf||_2^2$}
\STATE $\Dbf = \diag(\dbf_q)$.
\ENDIF
\ENDIF
\ENDWHILE
\end{algorithmic}
\end{algorithm}

\subsection{Complexity Analysis}
The complexity of the proposed RCG algorithm mainly lies in the iterative calculation of (\ref{equ.objFuncOriv3a}) and (\ref{equ.EucGrad}), of which the complexity is $\Ocal(M_tN_{\rm RF}M_{ris})$. Denote $I_r$ as the number of iterations of the RCG algorithm and denote $I_a$ as the average number of Armijo-Goldstein condition searches. Then the complexity of the RCG algorithm can be denoted as $T_{rcg} = \Ocal\left(I_r(1+I_a)M_tN_{\rm RF}M_{ris}\right)$, which grows linearly with the number of RIS elements.
For Algorithm \ref{Algo.1} that obtains $\Fbf_{\rm RF}$, the computational complexity of one inner iteration mainly lies in (\ref{equ.nablagh}), (\ref{equ.nablah2}), $\bar{\Qbf}_{m,k}$, and the TCG method, of which the complexity can be approximated as $\Ocal\left(M_r^3+I_cM_r^2\right)$ with $I_c$ being the iteration number of the TCG method.
Given the number of outer loop and inner loop are $N_{\rm RF}-1$ and $I_s$, the complexity of Algorithm \ref{Algo.1} is $T_1 = \Ocal\left((N_{\rm RF}-1)I_s(M_r^3+I_cM_r^2)\right)$. Hence the computational complexity of the Algorithm \ref{Algo.2} is $T_{tot} = I_d\left(T_{rcg}+T_1\right)$, where $I_d$ is the iteration number of Algorithm \ref{Algo.2}.

\section{Simulation Results}

\begin{figure}[htb]
\centering
{\psfig{figure=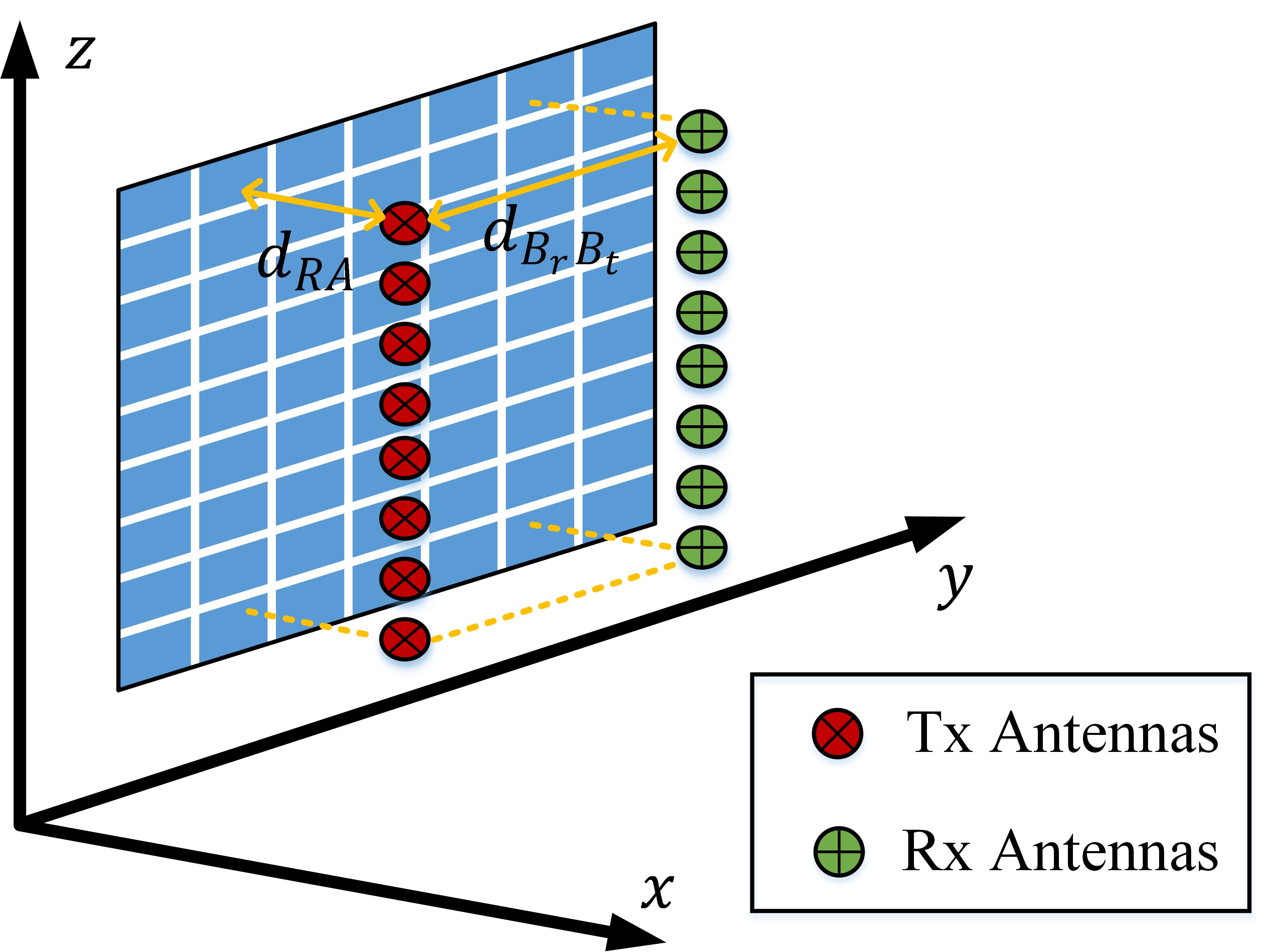,width= 2.5in}}
\caption{The placement of the RIS and the antennas.}
\label{fig.RABSRISULA}
\end{figure}

For the following simulations, the RIS is placed behind the antenna plane as illustrated in Fig. \ref{fig.RABSRISULA}, where both the Tx and Rx antennas are arranged into $8\times 1$ uniform linear array (ULA), i.e., $M_r=M_t=8$. The number of Rx RF chains is $N_{\rm RF}=3$. For both the antenna arrays and the RIS, the inter-element spacing is $\frac{\lambda}{2}$, where $\lambda$ is the wavelength.
Denote the distance between antennas and the RIS as $d_{RA}$, and the distance between the center of the transmit and receive antenna arrays as $d_{B_rB_t}$. We set $d_{B_rB_t} = 3\lambda$ and
$d_{RA} = \frac{\lambda}{2}$ \cite{10153667}.
For wave number $k_{\lambda}=\frac{2\pi}{\lambda}$, the LOS near-field channel between two adjacent points $A(x_1,y_1,z_1)$ and $B(x_2,y_2,z_2)$ in Fig. \ref{fig.RABSRISULA} is simulated as \cite{1552223}
\ben
h_{AB} = \sqrt{\beta_{AB}}e^{-jk_{\lambda}d_{AB}},
\label{equ.nearFieldChan}
\een
where $\beta_{AB} = \frac{G_l}{4}\left(\frac{1}{(k_{\lambda}d_{AB})^2}-\frac{1}{(k_{\lambda}d_{AB})^4}+\frac{1}{(k_{\lambda}d_{AB})^6}\right)$, $G_l=1$ given omnidirectional antenna, and $$d_{AB} = \sqrt{(x_1-x_2)^2+(y_1-y_2)^2+(z_1-z_2)^2}.$$
Hence the entries of $\Hbf_{B_rR}$, $\Hbf_{B_rB_t}$, and $\Hbf_{RB_t}$ are so generated from (\ref{equ.nearFieldChan}). For all the following simulations, it is assumed that $\lambda = 0.125$m.

Consider that the IBFD transceiver shown in Fig. \ref{fig.RABSRISULA} serves as a macro BS and communicates with three uplink (UL) users and three downlink (DL) users simultaneously. Both the UL and the DL users are of single-antenna and are uniformly distributed $[100{\rm m}, 140{\rm m}]$ away from the BS \cite{9318531}; the channels from users to the BS, from users to the RIS are from Saleh-Valenzuela model with free space path loss; the Tx power of the BS is $30$dBm, which is allocated with the well-known water-filling method and the zero-forcing precoding; the Tx power of UL users is $10$dBm and the thermal noise of the BS and the DL users are $-96$dBm \cite{8845768}. For Algorithm \ref{Algo.1}, we set $\xi = 0.6,\Omega_1=0.5$.

\begin{figure}[htb]
\centering
{\psfig{figure=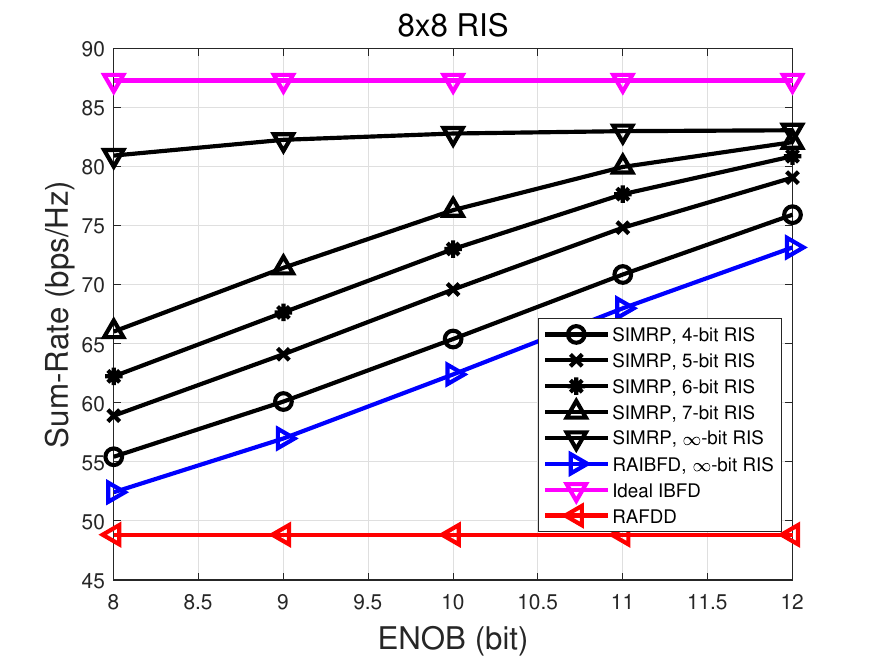,width= 3.2in}}
\caption{The sum rate vs. the effective number of bits.}
\label{fig.sumratevsenob}
\end{figure}
Given a $8\times 8$ RIS, Fig. \ref{fig.sumratevsenob} compares the sum rate performance of the SIMRP with that of the RAIBFD, ideal IBFD proposed in \cite{10153667} and RIS-assisted FDD (RAFDD) system, where the ideal IBFD system equipped with  $\infty$-bit ADCs and the RAFDD system provides an upper bound and a lower bound for the sum rate performance of the proposed SIMRP. Given $\infty$-bit RIS, the sum rate of the proposed SIMRP system reaches $95\%$ of that of the ideal IBFD as ADC's effective number of bits (ENOB) varies from $8$ to $12$, and outperforms the RAIBFD by around $33\%$ at ENOB $= 10$. Even when using RIS of finite bit resolutions, i.e., $b=4,5,6,7$, the SIMRP still has a gain of sum rate over the RAIBFD with $\infty$-bit RIS and the RAFDD system, but the RISs of lower bit resolution lead to larger performance degradation compared with the SIMRP with $\infty$-bit RIS, especially when $8$-bit and $9$-bit ADC are deployed. As $\infty$-bit PSN is considered in Fig. \ref{fig.sumratevsenob}, we can utilize PSN of high resolutions to approach its performance, i.e, $8$-bit PSN.

\begin{figure}[htb]
\centering
{\psfig{figure=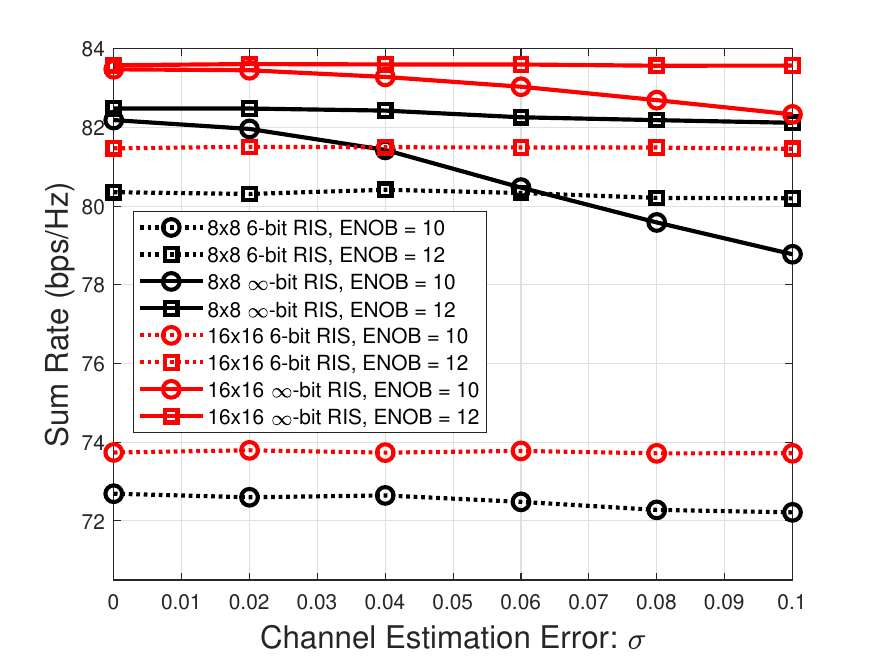,width= 3.2in}}
\caption{The sum rate vs. channel estimation error.}
\label{fig.sumratevscee}
\end{figure}

The second simulation studies the sum rate performance of the proposed SIMRP given imperfect channel state information (CSI), where the cascaded channels $\hat{\Hbf}_{B_rR}$ and $\hat{\Hbf}_{RB_t}$ are denoted as
\ben\label{Hest}
\begin{split}
\hat{\Hbf}_{B_rR} &= \Hbf_{B_rR} + {\Delta\Hbf_{B_rR}}\in{\mathbb C}^{M_r\times M_{ris}}, \\
\hat{\Hbf}_{RB_t} &= \Hbf_{RB_t} + {\Delta\Hbf_{RB_t}}\in{\mathbb C}^{M_{ris}\times M_{t}}.
\end{split}
\een
with the entries of ${\Delta\Hbf_{B_rR}}$ and ${\Delta\Hbf_{RB_t}}$ following ${\cal CN}\left(0,\frac{||\Hbf_{B_rR}||_F^2}{M_r^2M_{ris}^2}\sigma^2\right)$ and ${\cal CN}\left(0,\frac{||\Hbf_{RB_t}||_F^2}{M_{ris}^2M_t^2}\sigma^2\right)$. Fig. \ref{fig.sumratevscee} shows the sum rate performance of the SIMRP under different RIS sizes ($8\times8$ and $16\times16$) and different RIS resolutions ($6$-bit and $\infty$-bit) given that ENOB $= 10,12$, which shows that the SIMRP with RIS of larger size and ADC of reasonably high bit resolutions are robust to the channel estimation error.

\section{Conclusion} \label{SEC6}
In this paper, we propose a novel method, called self-interference mitigation using RIS and phase shifter network (SIMRP), where a co-located RIS and an analog phase shifter network (PSN) are utilized to mitigate the self-interference before the ADCs. We proposed to minimize the power of self-interference by jointly optimizing the phases of the RIS and the PSN. The simulation results show the proposed SIMRP method can effectively mitigate self-interference and outperform the state-of-the-art method.

\bibliographystyle{IEEEtran}
\bibliography{bib}
\balance
\end{document}

%% file: com.tex
\newcommand{\fs}{\hspace{0.07in}}
\newcommand{\bs}{\hspace{-0.1in}}
\newcommand{\re}{{\rm Re} \, }
\newcommand{\e}{{\rm E} \, }
\newcommand{\p}{{\rm P} \, }
\newcommand{\cn}{{\cal CN} \, }
\newcommand{\n}{{\cal N} \, }
\newcommand{\ba}{\begin{array}}
\newcommand{\ea}{\end{array}}
\newcommand{\be}{\begin{displaymath}}
\newcommand{\ee}{\end{displaymath}}
\newcommand{\ben}{\begin{equation}}
\newcommand{\een}{\end{equation}}
\newcommand{\bena}{\begin{eqnarray}}
\newcommand{\eena}{\end{eqnarray}}
\newcommand{\beqa}{\begin{eqnarray*}}
\newcommand{\enqa}{\end{eqnarray*}}
\newcommand{\f}{\frac}
\newcommand{\bc}{\begin{center}}
\newcommand{\ec}{\end{center}}
\newcommand{\bi}{\begin{itemize}}
\newcommand{\ei}{\end{itemize}}
\newcommand{\benu}{\begin{enumerate}}
\newcommand{\eenu}{\end{enumerate}}
\newcommand{\bdes}{\begin{description}}
\newcommand{\edes}{\end{description}}
\newcommand{\bt}{\begin{tabular}}
\newcommand{\et}{\end{tabular}}
\newcommand{\vs}{\vspace}
\newcommand{\hs}{\hspace}
\newcommand{\sort}{\rm sort \,}

\newcommand \thetabf{{\mbox{\boldmath$\theta$\unboldmath}}}
\newcommand{\Phibf}{\mbox{${\bf \Phi}$}}
\newcommand{\Psibf}{\mbox{${\bf \Psi}$}}
\newcommand \alphabf{\mbox{\boldmath$\alpha$\unboldmath}}
\newcommand \betabf{\mbox{\boldmath$\beta$\unboldmath}}
\newcommand \gammabf{\mbox{\boldmath$\gamma$\unboldmath}}
\newcommand \deltabf{\mbox{\boldmath$\delta$\unboldmath}}
\newcommand \epsilonbf{\mbox{\boldmath$\epsilon$\unboldmath}}
\newcommand \zetabf{\mbox{\boldmath$\zeta$\unboldmath}}
\newcommand \etabf{\mbox{\boldmath$\eta$\unboldmath}}
\newcommand \iotabf{\mbox{\boldmath$\iota$\unboldmath}}
\newcommand \kappabf{\mbox{\boldmath$\kappa$\unboldmath}}
\newcommand \lambdabf{\mbox{\boldmath$\lambda$\unboldmath}}
\newcommand \mubf{\mbox{\boldmath$\mu$\unboldmath}}
\newcommand \nubf{\mbox{\boldmath$\nu$\unboldmath}}
\newcommand \xibf{\mbox{\boldmath$\xi$\unboldmath}}
\newcommand \pibf{\mbox{\boldmath$\pi$\unboldmath}}
\newcommand \rhobf{\mbox{\boldmath$\rho$\unboldmath}}
\newcommand \sigmabf{\mbox{\boldmath$\sigma$\unboldmath}}
\newcommand \taubf{\mbox{\boldmath$\tau$\unboldmath}}
\newcommand \upsilonbf{\mbox{\boldmath$\upsilon$\unboldmath}}
\newcommand \phibf{\mbox{\boldmath$\phi$\unboldmath}}
\newcommand \varphibf{\mbox{\boldmath$\varphi$\unboldmath}}
\newcommand \chibf{\mbox{\boldmath$\chi$\unboldmath}}
\newcommand \psibf{\mbox{\boldmath$\psi$\unboldmath}}
\newcommand \omegabf{\mbox{\boldmath$\omega$\unboldmath}}
\newcommand \Sigmabf{\hbox{$\bf \Sigma$}}
\newcommand \Upsilonbf{\hbox{$\bf \Upsilon$}}
\newcommand \Omegabf{\hbox{$\bf \Omega$}}
\newcommand \Deltabf{\hbox{$\bf \Delta$}}
\newcommand \Gammabf{\hbox{$\bf \Gamma$}}
\newcommand \Thetabf{\hbox{$\bf \Theta$}}
\newcommand \Lambdabf{\hbox{$\bf \Lambda$}}
\newcommand \Xibf{\hbox{\bf$\Xi$}}
\newcommand \Pibf{\hbox{\bf$\Pi$}}
\newcommand \abf{{\bf a}}
\newcommand \bbf{{\bf b}}
\newcommand \cbf{{\bf c}}
\newcommand \dbf{{\bf d}}
\newcommand \ebf{{\bf e}}
\newcommand \fbf{{\bf f}}
\newcommand \gbf{{\bf g}}
\newcommand \hbf{{\bf h}}
\newcommand \ibf{{\bf i}}
\newcommand \jbf{{\bf j}}
\newcommand \kbf{{\bf k}}
\newcommand \lbf{{\bf l}}
\newcommand \mbf{{\bf m}}
\newcommand \nbf{{\bf n}}
\newcommand \obf{{\bf o}}
\newcommand \pbf{{\bf p}}
\newcommand \qbf{{\bf q}}
\newcommand \rbf{{\bf r}}
\newcommand \sbf{{\bf s}}
\newcommand \tbf{{\bf t}}
\newcommand \ubf{{\bf u}}
\newcommand \vbf{{\bf v}}
\newcommand \wbf{{\bf w}}
\newcommand \xbf{{\bf x}}
\newcommand \ybf{{\bf y}}
\newcommand \zbf{{\bf z}}
\newcommand \rbfa{{\bf r}}
\newcommand \xbfa{{\bf x}}
\newcommand \ybfa{{\bf y}}
\newcommand \Abf{{\bf A}}
\newcommand \Bbf{{\bf B}}
\newcommand \Cbf{{\bf C}}
\newcommand \Dbf{{\bf D}}
\newcommand \Ebf{{\bf E}}
\newcommand \Fbf{{\bf F}}
\newcommand \Gbf{{\bf G}}
\newcommand \Hbf{{\bf H}}
\newcommand \Ibf{{\bf I}}
\newcommand \Jbf{{\bf J}}
\newcommand \Kbf{{\bf K}}
\newcommand \Lbf{{\bf L}}
\newcommand \Mbf{{\bf M}}
\newcommand \Nbf{{\bf N}}
\newcommand \Obf{{\bf O}}
\newcommand \Pbf{{\bf P}}
\newcommand \Qbf{{\bf Q}}
\newcommand \Rbf{{\bf R}}
\newcommand \Sbf{{\bf S}}
\newcommand \Tbf{{\bf T}}
\newcommand \Ubf{{\bf U}}
\newcommand \Vbf{{\bf V}}
\newcommand \Wbf{{\bf W}}
\newcommand \Xbf{{\bf X}}
\newcommand \Ybf{{\bf Y}}
\newcommand \Zbf{{\bf Z}}
\newcommand \Omegabbf{{\bf \Omega}}
\newcommand \Rssbf{{\bf R_{ss}}}
\newcommand \Ryybf{{\bf R_{yy}}}
\newcommand \Cset{{\cal C}}
\newcommand \Rset{{\cal R}}
\newcommand \Zset{{\cal Z}}
\newcommand{\otheta}{\stackrel{\circ}{\theta}}
\newcommand{\defeq}{\stackrel{\bigtriangleup}{=}}
\newcommand{\oabf}{{\bf \breve{a}}}
\newcommand{\odbf}{{\bf \breve{d}}}
\newcommand{\oDbf}{{\bf \breve{D}}}
\newcommand{\oAbf}{{\bf \breve{A}}}
\renewcommand \vec{{\mbox{vec}}}
\newcommand{\Acalbf}{\bf {\cal A}}
\newcommand{\calZbf}{\mbox{\boldmath $\cal Z$}}
\newcommand{\feop}{\hfill \rule{2mm}{2mm} \\}
\newtheorem{theorem}{Theorem}[section]

\newcommand{\Rnum}{{\mathbb R}}
\newcommand{\Cnum}{{\mathbb C}}
\newcommand{\Znum}{{\mathbb Z}}
\newcommand{\Enum}{{\mathbb E}}

\newcommand{\Pcal}{{\cal P}}
\newcommand{\Lcal}{{\cal L}}
\newcommand{\Fcal}{{\cal F}}
\newcommand{\Scal}{{\cal S}}
\newcommand{\Qcal}{{\cal Q}}
\newcommand{\Mcal}{{\cal M}}
\newcommand{\Ccal}{{\cal C}}
\newcommand{\Dcal}{{\cal D}}
\newcommand{\Hcal}{{\cal H}}
\newcommand{\Ocal}{{\cal O}}
\newcommand{\Rcal}{{\cal R}}
\newcommand{\Zcal}{{\cal Z}}
\newcommand{\Xcal}{{\cal X}}
\newcommand{\zzbf}{{\bf 0}}
\newcommand{\zebf}{{\bf 0}}

\newcommand{\eop}{\hfill $\Box$}

\newcommand{\gss}{\mathop{}\limits}
\newcommand{\gs}{\mathop{\gss_<^>}\limits}

\newcommand{\circlambda}{\mbox{$\Lambda$
             \kern-.85em\raise1.5ex
             \hbox{$\scriptstyle{\circ}$}}\,}

\newcommand{\tr}{\mathop{\rm tr}}
\newcommand{\var}{\mathop{\rm var}}
\newcommand{\cov}{\mathop{\rm cov}}
\newcommand{\diag}{\mathop{\rm diag}}
\def\rank{\mathop{\rm rank}\nolimits}
\newcommand{\ra}{\rightarrow}
\newcommand{\ul}{\underline}
\def\Pr{\mathop{\rm Pr}}
\def\Re{\mathop{\rm Re}}
\def\Im{\mathop{\rm Im}}

\def\submbox#1{_{\mbox{\footnotesize #1}}}
\def\supmbox#1{^{\mbox{\footnotesize #1}}}

%
\newtheorem{Theorem}{Theorem}[section]
\newtheorem{Definition}[Theorem]{Definition}
\newtheorem{Proposition}[Theorem]{Proposition}
\newtheorem{Lemma}[Theorem]{Lemma}
\newtheorem{Corollary}[Theorem]{Corollary}
%
%
\newcommand{\ThmRef}[1]{\ref{thm:#1}}
\newcommand{\ThmLabel}[1]{\label{thm:#1}}
\newcommand{\DefRef}[1]{\ref{def:#1}}
\newcommand{\DefLabel}[1]{\label{def:#1}}
\newcommand{\PropRef}[1]{\ref{prop:#1}}
\newcommand{\PropLabel}[1]{\label{prop:#1}}
\newcommand{\LemRef}[1]{\ref{lem:#1}}
\newcommand{\LemLabel}[1]{\label{lem:#1}}
%

\newcommand \bbs{{\boldsymbol b}}
\newcommand \cbs{{\boldsymbol c}}
\newcommand \dbs{{\boldsymbol d}}
\newcommand \ebs{{\boldsymbol e}}
\newcommand \fbs{{\boldsymbol f}}
\newcommand \gbs{{\boldsymbol g}}
\newcommand \hbs{{\boldsymbol h}}
\newcommand \ibs{{\boldsymbol i}}
\newcommand \jbs{{\boldsymbol j}}
\newcommand \kbs{{\boldsymbol k}}
\newcommand \lbs{{\boldsymbol l}}
\newcommand \mbs{{\boldsymbol m}}
\newcommand \nbs{{\boldsymbol n}}
\newcommand \obs{{\boldsymbol o}}
\newcommand \pbs{{\boldsymbol p}}
\newcommand \qbs{{\boldsymbol q}}
\newcommand \rbs{{\boldsymbol r}}
\newcommand \sbs{{\boldsymbol s}}
\newcommand \tbs{{\boldsymbol t}}
\newcommand \ubs{{\boldsymbol u}}
\newcommand \vbs{{\boldsymbol v}}
\newcommand \wbs{{\boldsymbol w}}
\newcommand \xbs{{\boldsymbol x}}
\newcommand \ybs{{\boldsymbol y}}
\newcommand \zbs{{\boldsymbol z}}

\newcommand \Bbs{{\boldsymbol B}}
\newcommand \Cbs{{\boldsymbol C}}
\newcommand \Dbs{{\boldsymbol D}}
\newcommand \Ebs{{\boldsymbol E}}
\newcommand \Fbs{{\boldsymbol F}}
\newcommand \Gbs{{\boldsymbol G}}
\newcommand \Hbs{{\boldsymbol H}}
\newcommand \Ibs{{\boldsymbol I}}
\newcommand \Jbs{{\boldsymbol J}}
\newcommand \Kbs{{\boldsymbol K}}
\newcommand \Lbs{{\boldsymbol L}}
\newcommand \Mbs{{\boldsymbol M}}
\newcommand \Nbs{{\boldsymbol N}}
\newcommand \Obs{{\boldsymbol O}}
\newcommand \Pbs{{\boldsymbol P}}
\newcommand \Qbs{{\boldsymbol Q}}
\newcommand \Rbs{{\boldsymbol R}}
\newcommand \Sbs{{\boldsymbol S}}
\newcommand \Tbs{{\boldsymbol T}}
\newcommand \Ubs{{\boldsymbol U}}
\newcommand \Vbs{{\boldsymbol V}}
\newcommand \Wbs{{\boldsymbol W}}
\newcommand \Xbs{{\boldsymbol X}}
\newcommand \Ybs{{\boldsymbol Y}}
\newcommand \Zbs{{\boldsymbol Z}}

\newcommand \Absolute[1]{\left\lvert #1 \right\rvert}